\documentclass[cits]{PoS}
\pdfoutput=1
 
\usepackage{amsfonts,amsmath,amssymb,times,graphicx}
\usepackage{algorithmic}
\usepackage{algorithm}

\newcommand{\CC}{{\mathbb{C}}}
\newcommand{\kmax}{k_\text{max}}
\DeclareMathOperator{\sign}{sign}
\DeclareMathOperator{\spann}{span}
\DeclareMathOperator{\range}{range}
\newcounter{theorem}

\newtheorem{alg}[theorem]{Algorithm}

\title{Krylov subspace methods and the sign function: multishifts and
  deflation in the non-Hermitian case\thanks{Supported by the DFG
    collaborative research center SFB/TR-55 ``Hadron Physics from
    Lattice QCD''.}}

\ShortTitle{Krylov subspace methods and the sign function}

\author{Jacques C.R. Bloch$^a$, Tobias Breu$^a$, Andreas
  Frommer$^b$, Simon Heybrock$^a$, \speaker{Katrin Sch\"afer}$^b$,
  Tilo Wettig$^a$\\ 
  \\
  \llap{$^a$}Institute for Theoretical Physics, University of
  Regensburg, 93040 Regensburg, Germany\\
  \llap{$^b$}Department of Mathematics, University of Wuppertal,
  42097 Wuppertal, Germany\\

  E-mail: \\
  \email{jacques.bloch@physik.uni-regensburg.de},
  \email{tobias.breu@physik.uni-regensburg.de},
  \email{frommer@math.uni-wuppertal.de},
  \email{simon.heybrock@physik.uni-regensburg.de},
  \email{schaefer@math.uni-wuppertal.de},
  \email{tilo.wettig@physik.uni-regensburg.de} }

\abstract{Rational approximations of the matrix sign function lead to
  multishift methods. For non-Hermitian matrices long recurrences can
  cause storage problems, which can be circumvented with
  restarts. Together with deflation we obtain efficient iterative
  methods, as we show in numerical experiments for the overlap Dirac
  operator at non-vanishing quark chemical potential for lattices up
  to size $10^4$.}

\FullConference{The XXVII International Symposium on Lattice Field
  Theory\\
  July 26-31, 2009\\
  Peking University, Beijing, China}

\begin{document}

\section{Introduction}

In this paper we discuss the approximation of $f(A)b$, where $A \in
\CC^{n\times n}$ is non-Hermitian and $f$ is a function defined
on the spectrum of $A$ such that the extension of $f$ to matrix
arguments is defined.\footnote{The function $f$ can be extended to
  matrix arguments by, e.g., a spectral definition or a contour
  integral. For a thorough treatment of matrix functions see
  \cite{higham08}; a compact overview is given in \cite{frommer06}.}

The motivation for this rather general setting comes from quantum
chromodynamics (QCD) formulated on a discrete space-time lattice,
where $f = \sign$ is of special interest. As the main object relevant
for our discussion we are focusing on the overlap Dirac operator
\cite{Narayanan:1994gw,Neuberger:1997fp}. The main numerical effort
lies in the inversion of the overlap operator, which is done by
iterative methods and requires the repeated application of the sign
function of the usual ``symmetrized'' Wilson operator $H_W = \gamma_5
D_W$ (see \cite{Bloch:2006cd} for the notation) on
a vector.

At zero quark chemical potential $\mu$, the operator $H_W$ is
Hermitian. However, one can also study QCD at nonzero $\mu$, which is
relevant for many physical systems such as neutron stars, relativistic
heavy ion collisions, or the physics of the early universe. The
overlap operator has been generalized to this case
\cite{Bloch:2006cd,Bloch:2007xi}. The computational challenge is the
fact that at non-zero chemical potential $H_W$ becomes non-Hermitian.

This contribution is organized as follows. In Section 2 we review
multishift methods which have proven to be successful in the Hermitian
($\mu=0$) case. We will point out the problems that occur when
applying these methods to the non-Hermitian ($\mu\ne0$) case. In
Sections 3 and 4 we present two procedures, restarts and deflation,
which --- especially when applied in combination --- make multishift
methods applicable to non-Hermitian matrices.  We present our
numerical results in Section 5, and conclusions are drawn in Section
6.

\section{Multishift methods}

First we recall some results for the Hermitian case, i.e., we
investigate the computation of $f(A)b$, where $A \in
\CC^{n\times n}$ is Hermitian.  If $A$ is large, $f(A)$ is too
costly to compute, while $f(A)b$ can still be obtained in an efficient
manner if $A$ is sparse.  Krylov subspace methods, i.e., methods that
approximate $f(A)b$ in a Krylov subspace $K_k(A,b) = $span$\{b, Ab,
\dots, A^{k-1}b\}$, are suitable for this task.  We distinguish
between two Krylov subspace approaches: direct projection and
multishift.

Direct projection methods compute the sign function for the projection
of $A$ onto $K_k(A,b)$ and lift the result back to the original space,
see \cite{higham08, vdV87}, or \cite{Bloch:2007aw,Bloch:2008gh} in the
context of QCD. These methods are not the topic of this paper, but we
will use them for comparison in our numerical results.

The idea of multishift methods is to approximate $f$ by a rational
function $g$, 
\begin{align}
\label{eq:g}
f(x) \approx g(x) = \sum\limits_{i=1}^s \frac{\omega_i}{x - \sigma_i}\,.
\end{align}
The systems
\begin{equation}\label{shiftedsystems:eq}
(A - \sigma_i I) x^{(i)} = b\,, \quad i = 1, \dots, s
\end{equation}
are treated with standard Krylov subspace methods such as the
conjugate gradient method (CG) or the minimal residual method
(MINRES), approximating $x^{(i)}$ by ${x_k}^{(i)}$ from a Krylov
subspace.  Since Krylov subspaces are shift invariant, i.e.,
$K_k(A-\sigma_i I,b) = K_k(A,b)$, the approximations ${x_k}^{(i)}$ can
be computed simultaneously using the same subspace for all systems.
The desired approximation is then obtained by combining the
approximations to the $s$ shifted systems
\begin{align}
f(A)b \approx x_k = \sum\limits_{i=1}^s \omega_i {x_k}^{(i)}\,.
\end{align}
The core of any such method is the computation of an appropriate basis
for the Krylov subspace. For Hermitian matrices an orthonormal basis
can be built with short recurrences using the Lanczos process.  These
short recurrences are essential for the efficiency of the approach.

Turning to non-Hermitian matrices, the computation of an orthogonal
basis now requires long recurrences and is usually summarized via the
Arnoldi relation
\begin{align}
A V_k = V_k H_k + h _{k+1,k} v_{k+1} e_k^T\,.
\end{align} 
Here, $V_k = \left[ v_1 | \dots | v_k\right] \in \CC^{n\times
  k}$ is the matrix which contains the computed basis vectors (the
Arnoldi vectors), $H_k = {V_k}^\dagger A V_k$ is the upper Hessenberg
matrix containing the recurrence coefficients $h_{i,j}$, and $e_k$
denotes the $k$-th unit vector of $\CC^k$.

For the rational approximation approach this means that the
short-recurrence methods CG and MINRES have to be replaced by
multishift versions of the corresponding long-recurrence methods,
i.e., the full orthogonalization method (FOM) \cite{simoncini03} and
the generalized minimal residual method (GMRES) \cite{frommer98},
respectively.

Long recurrences slow down computation and increase storage
requirements, and thus become inefficient or even infeasible if $k$,
the dimension of the Krylov subspace, becomes large. In this paper we
investigate restarts to circumvent this problem for non-Hermitian
matrices.

\section{Restarts}

FOM to solve $A x = b$ consists of the Arnoldi process to compute the
Arnoldi vectors $v_1, \dots, v_k$ as well as the upper Hessenberg
matrix $H_k = {V_k}^\dagger A V_k$ and of approximating $x \approx x_k
= \|b\|_2 V_k {H_k}^{-1} e_1$. The Arnoldi process applied to $A -
\sigma_i I$ instead of $A$ produces the same matrices $V_k$ with $H_k$
replaced by the shifted counterpart $H_k - \sigma_i I$. The $k$-th
approximation to $g(A)b$, with $g(x)$ defined in \eqref{eq:g}, is thus
given by $\|b\|_2 \sum_{i=1}^s V_k (H_k-\sigma_i I)^{-1} e_1$.

To prevent recurrences from becoming too long one can --- in this case
--- use a restart procedure. This means that one stops the Arnoldi
process after $\kmax$ iterations. At this point we have a,
possibly crude, approximation to $g(A) b$, and to allow for a restart
one now has to express the error of this approximation anew as the
action of a matrix function, $g_1(A) b_1$, say.

A crucial observation concerning multishifts is that for any $k$ the
individual residuals ${r_k}^{(i)} = b - (A - \sigma_i I) {x_k}^{(i)}$
of the FOM iterates ${x_k}^{(i)}$ are just scalar multiples of the
Arnoldi vector $v_{k+1}$, see, e.g., \cite{simoncini03,frommer03}, i.e.,
\begin{align}
{r_k}^{(i)} = {\rho_k}^{(i)} v_{k+1}\,, \quad i = 1, \dots, s
\end{align}
with collinearity factors ${\rho_k}^{(i)} \in \CC$.  The error
$\Delta_k = g(A) b - x_k$ of the multishift approximation at step $k$
can therefore be expressed as
\begin{align}
\Delta_k = g_1(A) b_1\,, \quad \text{where } g_1(t) = \sum_{i=1}^s
\frac{\omega_i {\rho_k}^{(i)}}{t - \sigma_i} \text{ and }b_1=v_{k+1}\,.
\end{align}
This allows for a simple restart at step $\kmax$ of the Arnoldi
process, with the new function $g_1$ again being rational with the
same poles as $g$. This restart process can also be regarded as
performing restarted FOM for each of the individual systems
$(A-\sigma_i I) x = b$, $i=1,\ldots,s$ (and combining the individual
iterates appropriately), the point being that, even after a restart,
we need only a single Krylov subspace for all $s$ systems, see
\cite{simoncini03}.

There also exists a restarted version of multishift GMRES, see
\cite{frommer98} for a detailed derivation.

\section{Deflation}

In \cite{Bloch:2007aw} two deflation approaches were proposed which
use eigensystem information, namely Schur vectors (Schur deflation) or
left and right eigenvectors (LR deflation) corresponding to some
``critical'' eigenvalues. Critical eigenvalues are those which are
close to a singularity of $f$.  If they are not reflected very
precisely in the Krylov subspace, we get a poor approximation. In case
of the sign function the critical eigenvalues are those close to the
imaginary axis.  Here, we describe LR deflation (see
\cite{Bloch:2009in} for the reason why this is the method of choice)
and show how it can be combined with multishifts and restarts.

Let $R_m = \left[r_1 | \dots | r_m \right]$ be the matrix containing
the right eigenvectors and ${L_m}^\dagger = \left[l_1 | \dots |
  l_m\right]^\dagger$ the matrix containing the left eigenvectors
corresponding to $m$ critical eigenvalues of the matrix $A$. This
means that we have 
\begin{align}
A R_m = R_m \Lambda_m \quad \text{and} \quad
{L_m}^\dagger A = \Lambda_m {L_m}^\dagger\,,
\end{align}
where $\Lambda_m$ is a diagonal matrix containing the $m$ critical
eigenvalues.  Since left and right eigenvectors are
biorthogonal, we can normalize them such that ${L_m}^\dagger R_m =
I_m$. 
The matrix $P = R_m {L_m}^\dagger$ represents an oblique projector
onto the subspace $\Omega_R = \spann\{r_1, \dots,r_m\}$.  

We now split $f(A)b$ into the two parts
\begin{align}
   f(A)b = f(A)(Pb) + f(A)(I-P)b\,.
\end{align}
Since we know the left and right eigenvectors which make up $P$, we
directly obtain
\begin{align}
 x_P \equiv f(A)(Pb) = f(A)R_m L_m^\dagger b = R_m f(\Lambda_m) (L_m^\dagger b)\,,
\end{align}
which can be computed exactly.  The remaining part $f(A)(I-P)b$ can 
then be approximated iteratively by using a multishift
method. Thus $f(A)b$ is now approximated in augmented Krylov subspaces
$\Omega_R + K_k(A,(I - P)b)$,
\begin{align}
  x_k = \underbrace{x_P\phantom{\Bigg|}\!\!\!\!}_{\in\Omega_R} 
  + \;\;\underbrace{\sum_{i=1}^s \omega_i x_k^{(i)}}_{\makebox[0mm]{\scriptsize$\in K_k(A,(I - P)b)$}}\,.
\end{align}

Theoretically, we have
\begin{align}
  K_k(A,(I-P)b) = (I-P)K_k(A,(I-P)b) \subseteq \range(I-P)\,,
\end{align}
see \cite{Bloch:2009in}. In computational practice, however, components outside of
$\range(I-P)$ will show up gradually when building $K_k(A,(I-P)b)$ due to rounding effects in
floating-point arithmetic. It is thus necessary to reapply $I-P$
from time to time in order to eliminate these components.

Since the only effect of LR deflation is the replacement of
$b$ by $(I-P)b$, no modifications of the restart algorithm are
necessary.

\section{Algorithms}

We combine multishift methods with restarts and deflation. We assume
that the original function $f$ is replaced by a rational function
(given by the shifts $\sigma_i$ and weights $\omega_i$) which
approximates the original function sufficiently well after deflation.

Depending on the underlying multishift method (FOM or GMRES), we get
LR-deflated multishift FOM (FOM-LR) or LR-deflated multishift GMRES
(GMRES-LR). Algorithm~\ref{fom:alg} gives an algorithmic description
of FOM-LR.  (For an algorithmic description of GMRES-LR we refer to
\cite{Bloch:2009in}.)  The notation FOM-LR($m,k$) indicates that we
LR-deflate a subspace of dimension $m$ and that we restart FOM after a
cycle of $k$ iterations.  The vector $x$ is the approximation to
$f(A)b$.  After the completion of each cycle we perform a projection
step to eliminate numerical contamination by components outside of
$\range(I-P)$.
 
\begin{algorithm}
\begin{alg}\rm Restarted FOM-LR($m,k$)\label{fom:alg}
\begin{algorithmic}
\STATE \{{\bf Input} $m$, $k=\kmax$, $A$, $\{\sigma_1, \dots, \sigma_s\}$, $\{\omega_1,\ldots,
\omega_s\}$, $b$, $L=L_m$, $R=R_m$, $\Lambda=\Lambda_m$\}
\STATE $x = x_P = R f(\Lambda) {L}^\dagger b$
\STATE $r = (I-P)b$
\STATE $\rho^{(i)} = 1$, $i = 1, \dots, s$
\WHILE[\emph{loop over restart cycles}]{not all systems are converged}
\STATE $\beta = \|r\|_2$
\STATE $v_1 = r/\beta$
\STATE compute $V_k$, $H_k$ by running $k$ steps of Arnoldi with $A$
\STATE $y_k^{(i)} = \beta \rho^{(i)} (H_k - \sigma_i I_k)^{-1} e_1$, $i = 1, \dots, s$
\STATE $x = x + V_k\sum_{i=1}^s \omega_i y_k^{(i)}$
\STATE $r = v_{k+1}$
\STATE $\rho ^{(i)} = - h_{k+1,k} e_k^T y_k^{(i)}$, $i = 1, \dots, s$
\STATE $r = (I-P)r$ \COMMENT{\emph{projection step}}
\ENDWHILE
\end{algorithmic}
\end{alg}
\end{algorithm} 

Note that a combination of deflation and a multishift method based on
the two-sided \mbox{Lanczos} algorithm is also possible, see
\cite{Bloch:2009in}. Of course, since two-sided Lanczos already
gives short recurrences, there is no need to restart here.

\section{Numerical results}

For our numerical experiments we turn to $f = \sign$. In the
Hermitian case, the sign function of $A$ can be approximated using the
Zolotarev best rational approximation, see \cite{zolotarev77} and,
e.g., \cite{ingerman00, vandenEshof:2002ms}. Using the Zolotarev
approximation on non-Hermitian matrices gives rather poor results,
unless all eigenvalues are close to the real axis.  A better choice
for generic non-Hermitian matrices is the rational approximation
originally suggested by Kenney and Laub \cite{KL} and used by
Neuberger \cite{Neuberger:1998my, Neuberger:1999zk} for vanishing
chemical potential,
\begin{align}
  \sign(t) \approx g_s(t)\,, \quad \text{where }  
   g_s(t) = \frac{(t+1)^{2s} - (t-1)^{2s}}{(t+1)^{2s}
    + (t-1)^{2s}}\,.
\end{align}
The partial fraction expansion of $g_s$ is known to be
\begin{align}
  g_s(t) = t \sum_{i=1}^s \frac{\omega_i}{t^2 - \sigma_i} \quad
  \text{with } 
  \omega_i = \frac{1}{s} \cos^{-2}\left( \frac{\pi}{2 s}
    \left(i-\frac{1}{2}\right)\right), \quad \sigma_i =
  -\tan^2\left(\frac{\pi}{2s}\left(i-\frac{1}{2}\right)\right), 
\end{align}
see \cite{KL,Neuberger:1998my}. Note that actually one uses $g(ct)$,
where the parameter $c>0$ is chosen to minimize the number of poles
$s$ needed to achieve a given accuracy. If the spectrum of $A$ is
known to be contained in the union of two circles $C(m,r) \cup
C(-m,r)$, where $C(m,r)$ is the circle $\{ |z-m| \leq r\}$ and $m$ and
$r$ are real with $0 < r <m$, then $c = ((m+r)(m-r))^{-1/2}$ is
optimal, see \cite{Bloch:2009in, vandenEshof:2002ms}.

\begin{figure}[h]
   \includegraphics[width=0.48\textwidth]{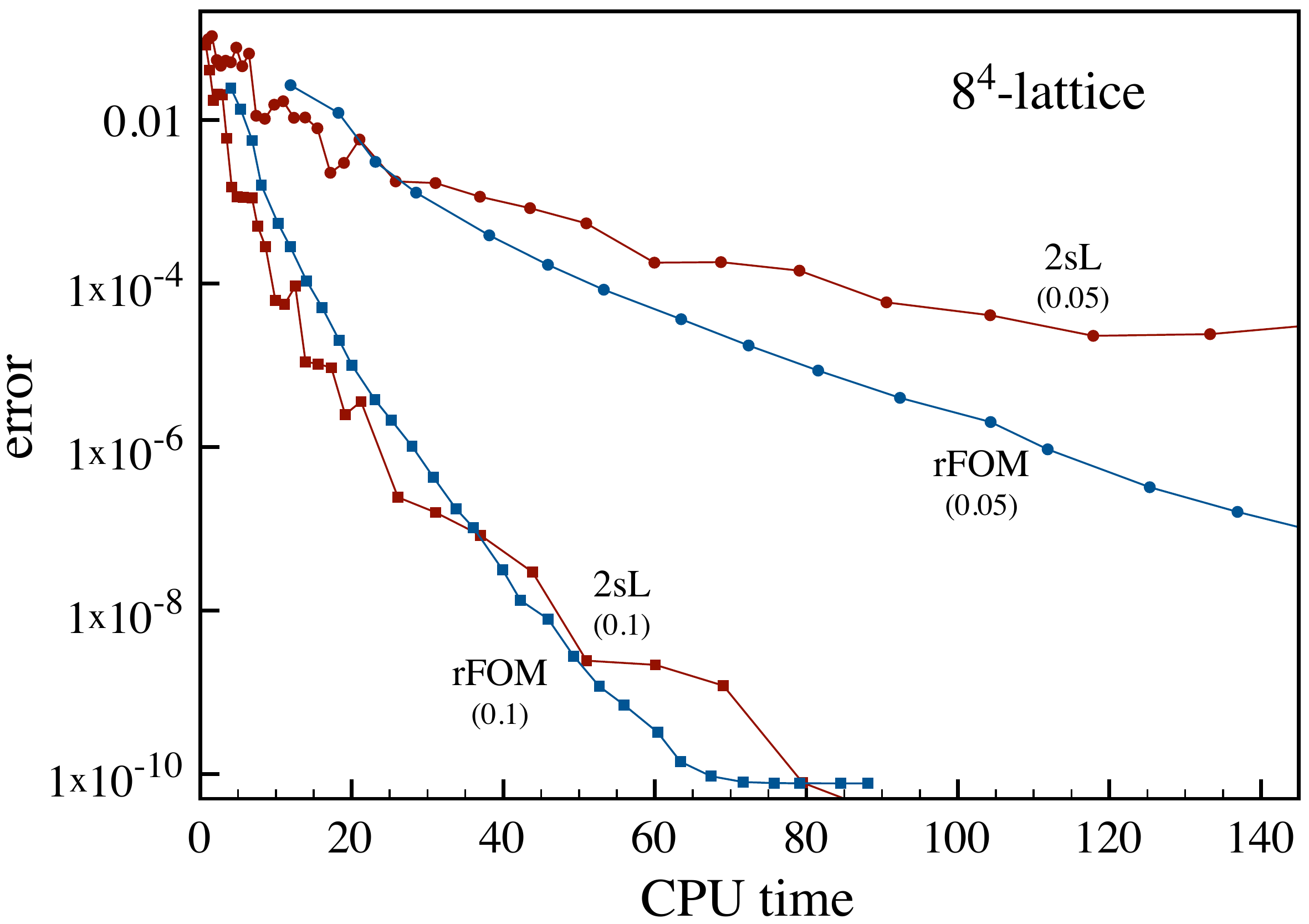}\hfill
   \includegraphics[width=0.48\textwidth]{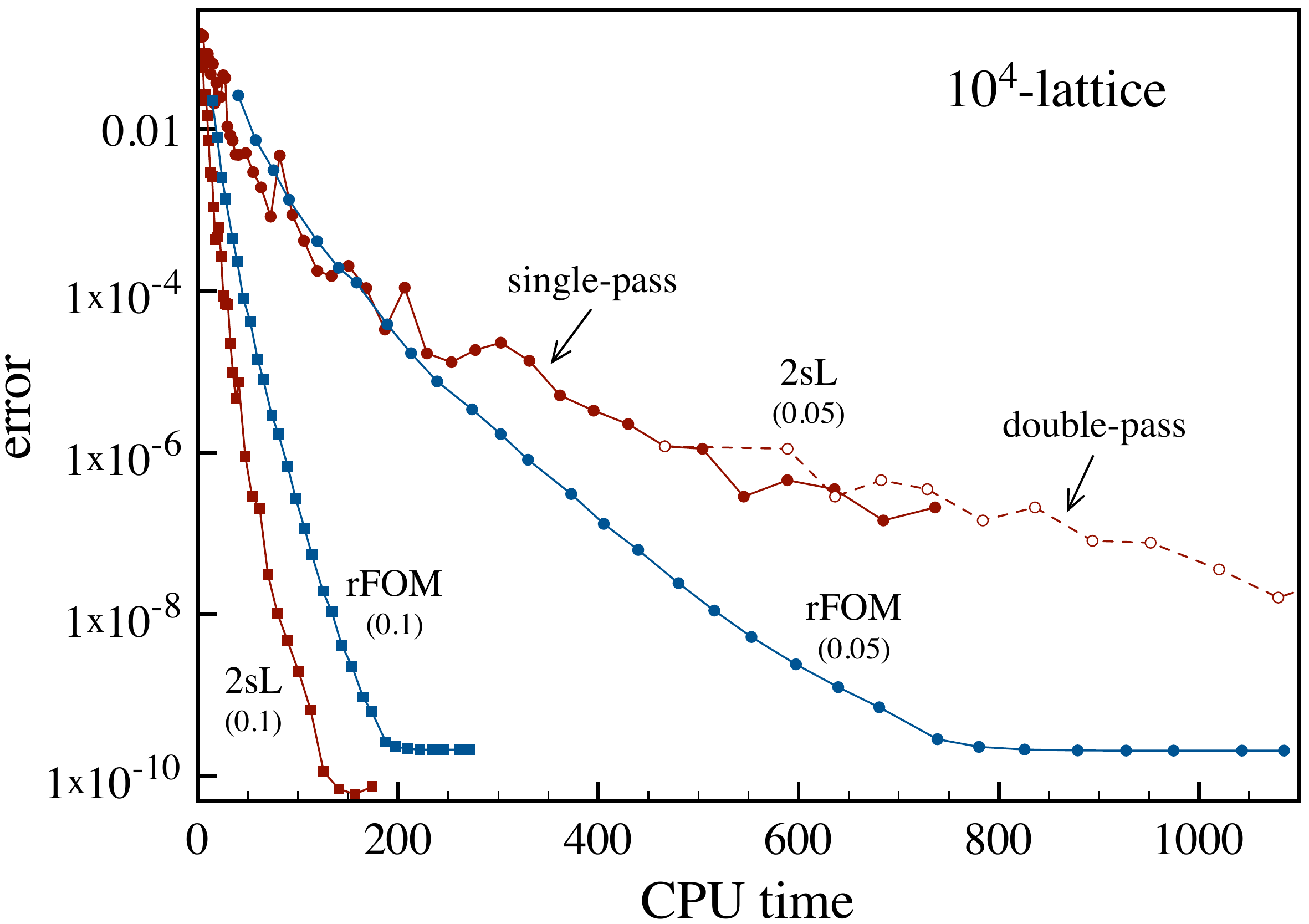}
 \caption{\label{Fig:error} Comparison of the accuracy of the
   restarted FOM-LR algorithm (rFOM) and the direct two-sided
   Lanczos-LR method (2sL) as a function of the CPU time in seconds
   for an $8^4$ (left) and a $10^4$ (right) lattice configuration,
   using $\mu=0.3$ in both cases.  Each plot shows data for two
   different deflation gaps, given in parentheses. The restart size
   used in the restarted FOM-LR algorithm is $\kmax = 30$ for the
   $8^4$ lattice and $\kmax = 40$ for the $10^4$ lattice.}
\end{figure}

Figure \ref{Fig:error} shows the performance of FOM-LR in comparison
to the direct projection method.  The $k$-th approximation in the
latter is given as
\begin{equation} \label{direct:eq}
   x_P+\| (I-P)b \|_2 V_k \sign(H_k) e_1\,,
\end{equation}
where $\sign(H_k)$ is computed via Roberts' method, see
\cite{higham08}.  The relative performance of the two approaches
depends on the parameters of the problem, such as the lattice size,
the deflation gap, and the size of the Krylov subspace.  For more
details, see \cite{Bloch:2009in}.  We add that in the meantime an
improved method to compute $\sign(H_k)$ in the direct approach has
been developed, see \cite{Bloch:2009pos} in these proceedings.

\pagebreak

\begin{figure}[h]
  \includegraphics[width=0.48\textwidth]{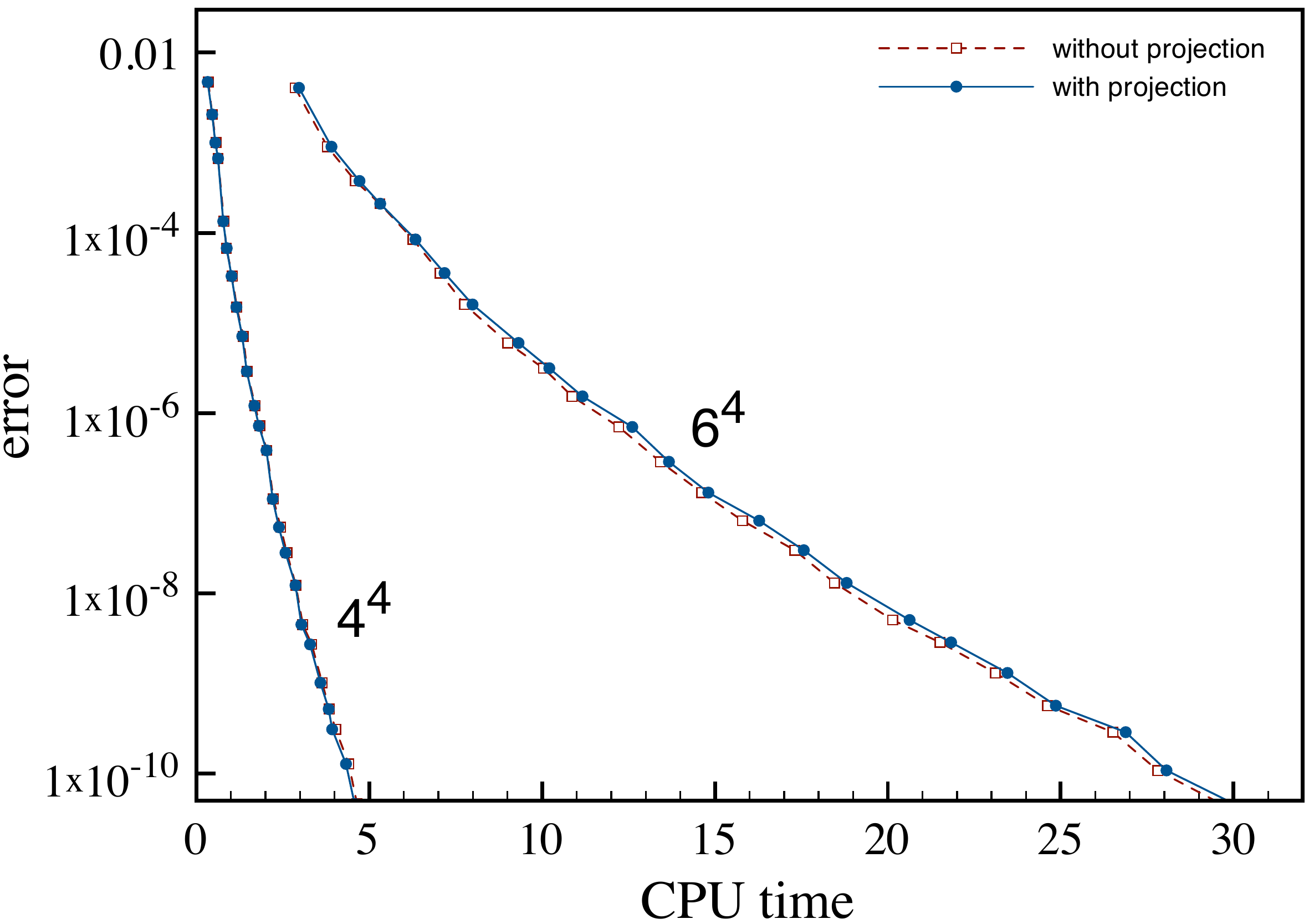}\hfill
  \includegraphics[width=0.48\textwidth]{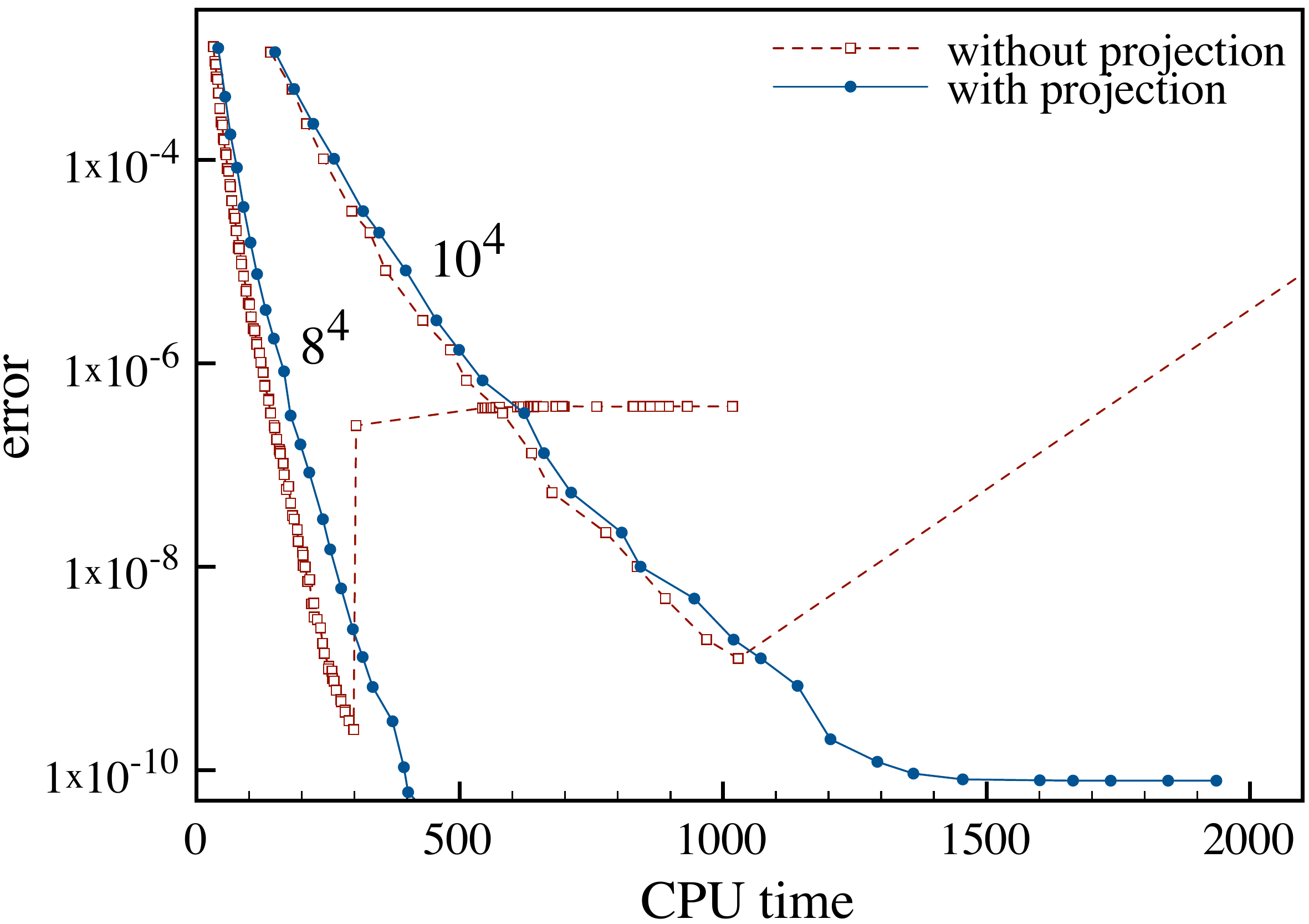}
\caption{\label{Fig:projection}Error vs CPU time for the FOM-LR
  algorithm with and without re-orthogonalization for $4^4$ and $6^4$
  (left) as well as $8^4$ and $10^4$ (right) lattices.  We again used
  $\mu=0.3$ in all cases.}
\end{figure}

Figure \ref{Fig:projection} is meant to convey a warning.  It shows
that the projection step after each restart, as formulated in
Algorithm~\ref{fom:alg}, may be crucial to ensure convergence. In both
plots we give results for Algorithm~\ref{fom:alg} and a variant
thereof in which the projection step is omitted. The right plot shows
that this may destroy convergence, the left plot shows that this is
not necessarily so.  Since the CPU time is increased only marginally
by the projection step, the latter should always be included.

\section{Conclusion}

We have presented an algorithm, FOM-LR, to approximate the action of
the sign function of a non-Hermitian matrix on a vector.  This
algorithm combines LR deflation and a rational approximation to the
sign function, which is computed by a restarted multishift method.
The latter has fixed storage requirements determined by the restart
parameter (maximum size of the Krylov subspace) and the degree of the
rational approximation.  Occasionally, additional projections of the
Krylov vectors are necessary for numerical stability.

Whether FOM-LR or a direct method (i.e., the two-sided Lanczos-LR
method) performs better depends on many details of the problem.  Some
of them have been mentioned in Section 6.  Others include
implementation issues such as optimized linear algebra libraries, and
ultimately parallelization.

\bibliographystyle{JHEP}

\bibliography{my_lit}

\end{document}